\documentclass[aps,prl,twocolumn,preprintnumbers,superscriptaddress,nofootinbib]{revtex4-1}
\usepackage[utf8]{inputenc}
\usepackage{graphicx}
\usepackage{psfrag}
\usepackage{amssymb}
\usepackage{amsmath}
\usepackage{epstopdf}
\usepackage{color}
\begin{document}
\title{Parton Distribution Functions from a Light Front Hamiltonian and\\ QCD Evolution for Light Mesons}
\author{Jiangshan Lan}
\email{jiangshanlan@impcas.ac.cn}
\affiliation{Institute of Modern Physics, Chinese Academy of Sciences, Lanzhou 730000, China}
\affiliation{University of Chinese Academy of Sciences, Beijing 100049, China}
\author{Chandan Mondal}
\email{mondal@impcas.ac.cn} 
\affiliation{Institute of Modern Physics, Chinese Academy of Sciences, Lanzhou 730000, China}
\author{Shaoyang Jia}
\email{sjia@iastate.edu} 
\affiliation{Department of Physics and Astronomy, Iowa State University, Ames, Iowa 50011, USA}
\author{Xingbo Zhao}
\email{xbzhao@impcas.ac.cn} 
\affiliation{Institute of Modern Physics, Chinese Academy of Sciences, Lanzhou 730000, China}
\affiliation{University of Chinese Academy of Sciences, Beijing 100049, China}
\author{James P. Vary}
\email{jvary@iastate.edu} 
\affiliation{Department of Physics and Astronomy, Iowa State University, Ames, Iowa 50011, USA}

\collaboration{BLFQ Collaboration}

\begin{abstract}
We obtain the pion and the kaon parton distribution functions from the eigenstates of a light front effective Hamiltonian in the constituent quark-antiquark representation suitable for low-momentum scale applications. By taking these scales as the only free parameters, the valence quark distribution functions of the pion, after QCD evolution, are consistent with the data from the FNAL-E615 experiment. The ratio of the up quark distribution of the kaon to that of the pion also agrees with the CERN-NA3 experiment. Supplemented by known parton distribution functions for the nucleons, we further obtain the cross section consistent with experimental data for the ${\pi^-\rm{nucleus}}\rightarrow{\mu^+\mu^-X}$ Drell-Yan process. 
\end{abstract}
\maketitle
\section{Introduction} 
The parton distribution function (PDF), the probability that a parton (quark or gluon) carries a certain fraction of the total light front momentum of a hadron, encodes the nonperturbative structure of hadrons, attracting numerous dedicated experimental and theoretical efforts \cite{Bordalo:1987cs,Freudenreich:1990mu,Sutton:1991ay,Gluck:1999xe,Wijesooriya:2005ir,Badier:1983mj, Conway:1989fs,Aicher:2010cb,Watanabe:2017pvl,Hecht:2000xa,Nam:2012vm,Detmold:2003tm,Holt:2010vj,Pumplin:2002vw,Ball:2017nwa,Alekhin:2017kpj,Dulat:2015mca,Harland-Lang:2014zoa,Barry:2018ort}. Since experimental results span a large range of momentum transfers, one must address the dependence of these results on the resolving power (scale) of the experimental probe, which is equivalent to addressing the physics of scale evolution based on quantum chromodynamics (QCD)~\cite{peskin1995introduction}. Starting with an effective Hamiltonian for a constituent quark and an antiquark (masses of several hundred $\mathrm{MeV}$), suitable for low-resolution probes, we solve for the light front wave functions (LFWFs) of the pion and the kaon to produce the initial PDFs. We then apply QCD evolution from the initial PDFs to account for the emission and absorption of sea quarks and gluons in order to incorporate degrees of freedom relevant to higher-resolution probes. This then allows us to compare our QCD-evolved PDFs with various sets of experimental data over a wide range of scales. 

Two salient issues can be addressed with this approach. The first issue is the valence PDF of the pion, which has been investigated in theory by Refs.~\cite{Broniowski:2007si,Gutsche:2013zia,Ahmady:2018muv,deTeramond:2018ecg,Brommel:2006zz,Martinelli:1987bh,Detmold:2003tm,Abdel-Rehim:2015owa,Oehm:2018jvm,Sufian:2019bol,Barry:2018ort}. Experimentally, this PDF is measured with the pion-nucleus-induced Drell-Yan (DY) process, in which a quark annihilates with an antiquark and produces a dilepton pair~\cite{Bordalo:1987cs,Freudenreich:1990mu,Sutton:1991ay,Gluck:1999xe,Wijesooriya:2005ir}. Specifically, there is a disagreement on the behavior of the pion valence PDF when either the quark or the antiquark approaches the limit of taking all of the pion's light front momentum (\textit{i.e.} the annihilating parton's light front momentum fraction $x$ approaches unity)~\cite{Farrar:1979aw,Berger:1979du,Brodsky:2006hj,Yuan:2003fs,Hecht:2000xa,Frederico:1994dx,Shigetani:1993dx,Melnitchouk:2002gh,Aicher:2010cb}. The second issue concerns the description of the experimental data on the kaon valence PDF, which exists in the form of the ratio of the up ($u$) quark distribution in the kaon to that in the pion~\cite{Badier:1983mj, Conway:1989fs}. The valence PDF of the kaon has been theoretically investigated in Refs.~\cite{Nguyen:2011jy,Shi:2018mcb,Watanabe:2017pvl,Nam:2012vm,Chen:2016sno,Hutauruk:2016sug,Davidson:2001cc,Xu:2018eii,Broniowski:2017wbr,Radyushkin:2017gjd}. Similar to the scale dependence of the angular momentum observables~\cite{Thomas:2008ga,Jia:2012wf}, addressing these two issues requires a unified approach, such as we describe here, that successfully encapsulates properties of both the pion and the kaon at their respective model scales, while the available data across various other scales are then modeled with reasonable precision after QCD evolution. 

With the theoretical framework of basis light front quantization (BLFQ)~\cite{Vary:2009gt,Wiecki:2014ola,Li:2015zda}, we adopt an effective light front Hamiltonian~\cite{Jia:2018ary} and solve for its mass eigenstates at the scales suitable for low-resolution probes. With quarks being the only explicit degrees of freedom for the strong interaction, our Hamiltonian incorporates the holographic QCD confinement potential~\cite{deTeramond:2018ecg} supplemented by the longitudinal confinement~\cite{Li:2017mlw}. Our Hamiltonian also includes the color-singlet Nambu--Jona-Lasinio (NJL) interactions~\cite{Klimt:1989pm,Shigetani:1993dx} to account for the dynamical chiral symmetry breaking of QCD. By solving this Hamiltonian in the constituent quark-antiquark Fock space (the valence space) and fitting the quark masses and coupling constants, one obtains via the pion and the kaon LFWFs the good quality descriptions of their charge radii, distribution amplitudes, and electromagnetic form factors~\cite{Jia:2018ary}. 

Here, we evaluate the pion and the kaon PDFs from their LFWFs obtained in Ref.~\cite{Jia:2018ary}, apply QCD evolution to higher momentum scales, and compare results with an experiment where available. We introduce independent initial low-momentum scales of the effective Hamiltonians for the pion and the kaon. These two scales serve as reference points where the meson structure is effectively described by the valence (constituent) quark and antiquark pair, and we take them as the only two adjustable parameters which we determine by requiring the evolved PDFs to fit a selection of data. Although our own approach is not perfect, it will be shown to provide a good description for a wide range of available data.
\section{Pion and kaon valence PDFs from BLFQ}
At the initial scale where the mesons are described by a quark and antiquark pair, we adopt the light front effective Hamiltonian $H_{\rm eff}$ defined by
\begin{align}
H_\mathrm{eff} =& \frac{\vec k^2_\perp + m_q^2}{x} + \frac{\vec k^2_\perp+m_{\bar q}^2}{1-x}
+ \kappa^4 x(1-x) {\vec r}_\perp^{\,2}\nonumber\\
&- \frac{\kappa^4}{(m_q+m_{\bar q})^2} \partial_x[x(1-x) \partial_x ]+H^{\rm eff}_{\rm NJL},\label{eqn:Heff}
\end{align}
where $m_q$ ($m_{\bar q}$) is the mass of the quark (antiquark) and $\kappa$ is the strength of the confinement. Meanwhile, $\vec{k}_\perp$ is the relative transverse momentum, while $\vec{r}_\perp$ is its conjugate related to the holographic variable~\cite{Brodsky:2014yha}. Additionally, $H^{\rm eff}_{\rm NJL}$ accounts for the chiral dynamics, which takes the form of the four-fermion contact interaction in the color-singlet NJL model~\cite{Klimt:1989pm}. Specifically, we adopt the NJL interactions in the scalar-pseudoscalar channel. In the valence Fock sector of the $\pi^+$, we obtain 
\begin{align}
 H_{\mathrm{NJL},\pi}^{\mathrm{eff}} & =G_\pi\, \big\{\overline{u}_{\mathrm{u}s1'}(p_1')u_{\mathrm{u}s1}(p_1)\,\overline{v}_{\mathrm{d}s2}(p_2)v_{\mathrm{d}s2'}(p_2') \nonumber\\
 &\quad+ \overline{u}_{\mathrm{u}s1'}(p_1')\gamma_5 u_{\mathrm{u}s1}(p_1)\,\overline{v}_{\mathrm{d}s2}(p_2)\gamma_5 v_{\mathrm{d}s2'}(p_2') \nonumber\\
 &\quad+ 2\,\overline{u}_{\mathrm{u}s1'}(p_1')\gamma_5 v_{\mathrm{d}s2'}(p_2')\,\overline{v}_{\mathrm{d}s2}(p_2)\gamma_5 u_{\mathrm{u}s1}(p_1) \big\},\label{eq:H_eff_NJL_pi_ori}
\end{align}
as the $H^{\mathrm{eff}}_{\mathrm{NJL}}$ term in Eq.~\eqref{eqn:Heff}. While in the valence Fock sector of $K^+$, we obtain 
\begin{align}
H^{\mathrm{eff}}_{\mathrm{NJL},K}&=G_K\,\big\{- 2\,\overline{u}_{\mathrm{u}s1'}(p_1') v_{\mathrm{s}s2'}(p_2')\,\overline{v}_{\mathrm{s}s2}(p_2) u_{\mathrm{u}s1}(p_1) \nonumber\\
&\quad + 2\,\overline{u}_{\mathrm{u}s1'}(p_1')\gamma_5 v_{\mathrm{s}s2'}(p_2')\,\overline{v}_{\mathrm{s}s2}(p_2)\gamma_5 u_{\mathrm{u}s1}(p_1) \big\},\label{eq:H_eff_NJL_SU_3_ori}
\end{align}
as the $H^{\mathrm{eff}}_{\mathrm{NJL}}$ term in Eq.~\eqref{eqn:Heff}. Here ${u_{\mathrm{f}s}(p)}$ and ${v_{\mathrm{f}s}(p)}$ are the solutions of the Dirac equation, with the nonitalic subscripts representing the flavors and the italic subscripts designating the spins. Meanwhile, $p_1$ and $p_2$ are the momenta of the valence quark and valence antiquark, respectively~\cite{Jia:2018ary}. 

Our current treatment of the BLFQ-NJL model is unable to reach the chiral limit. Specifically, the quark self-energy due to the NJL interactions has been modeled by the large valence quark masses for the light and strange flavors. Consequently, the NJL interactions in Eqs.~\eqref{eq:H_eff_NJL_pi_ori}~and~\eqref{eq:H_eff_NJL_SU_3_ori}, manifesting as the spin-orbital coupling in the meson systems, account for the relatively small masses of $\pi^+$ and $K^+$ compared to those of $\rho^+$ and $K^{*+}$.

The mass eigenstates $\vert\Psi\rangle$ are solutions of ${H_{\mathrm{eff}} \vert \Psi\rangle = M^2 \vert\Psi\rangle}$, where $M$ is the eigenmass. They are expressed as superpositions of a quark-antiquark pair in BLFQ modes of relative motion. The amplitudes of these superpositions constitute the valence LFWF for each state. We refer to our model as the BLFQ-NJL model. Parameters in our model are adjusted to reproduce the masses of the $\pi^+$, $\rho^+$, $K^+$, and $K^{*+}$, as well as the experimental charge radii of $\pi^+$ and $K^+$~\cite{Jia:2018ary}. 

The valence wave function in the momentum space is expanded in an orthonormal basis set designed to preserve the symmetries of the effective Hamiltonian:
\begin{align}
&\quad \psi_{rs}(x,\overrightarrow{k}^\perp)=\sum_{nml}\psi(n,m,l,r,s)\,\dfrac{4\pi}{\kappa}\sqrt{(2l+\alpha+\beta+1)}\nonumber\\
&\quad \times\sqrt{\dfrac{ n!}{(n+|m|)!} \dfrac{\Gamma(l+1)\Gamma(l+\alpha+\beta+1)}{\Gamma(l+\alpha+1)\Gamma(l+\beta+1)}} \nonumber\\
& \quad \times \left(\dfrac{\vert\overrightarrow{q}^\perp\vert}{\kappa}\right)^{|m|} \exp\left(-\dfrac{\overrightarrow{q}^{\perp 2}}{2\kappa^2}\right)
\,L_n^{|m|} \left(\dfrac{\overrightarrow{q}^{\perp 2}}{\kappa^2}\right)\,e^{im\varphi}\nonumber\\
& \quad \times x^{\beta/2}(1-x)^{\alpha/2}\,P_l^{(\alpha,\beta)}(2x-1),\label{eq:psi_rs_basis_expansions}
\end{align}
with $\overrightarrow{q}^\perp=\overrightarrow{k}^\perp/\sqrt{x(1-x)}$ and $\tan(\varphi)=k^2/k^1$. Here $L_n^{|m|}$ is the associated Laguerre function. The integer $n\,(m)$ is the radial (orbital angular momentum projection) quantum number of the two-dimensional (2D) harmonic oscillator (HO) function employed in Eq.~\eqref{eq:psi_rs_basis_expansions}. The quantity $P_{l}^{(\alpha,\beta)}(z)$ is the Jacobi polynomial with $l$ corresponding to the quantum number of longitudinal excitations, ${\alpha =2m_{\overline{q}}(m_q+m_{\overline{q}})/\kappa^2}$, and ${\beta=2m_q(m_q+m_{\overline{q}})/\kappa^2}$. The subscripts ``$r$" and ``$s$" are the spin labels for the valence quarks. 
Each term in Eq.~\eqref{eq:psi_rs_basis_expansions} is an eigenfunction for $H_{\mathrm{eff}}$ without $H_{\mathrm{NJL}}^{\mathrm{eff}}$. However, the Dirac structures of the NJL interactions in Eqs.~\eqref{eq:H_eff_NJL_pi_ori} and \eqref{eq:H_eff_NJL_SU_3_ori} result in nontrivial spin wave functions in the solutions obtained in the form of Eq.~\eqref{eq:psi_rs_basis_expansions}. 

Next, we truncate the infinite basis by restricting ${0 \leq n \leq N_{\mathrm{max}}=8}$, ${-2 \leq m \leq 2}$, and ${0 \leq l \leq L_{\mathrm{max}}}$ and diagonalize Eq.~\eqref{eqn:Heff} numerically in the representation of Eq.~\eqref{eq:psi_rs_basis_expansions} to obtain the LFWFs~\cite{Jia:2018ary}. Here $L_{\text{max}}$ is the basis resolution in the longitudinal direction, whereas $N_{\text{max}}$ controls the transverse momentum covered by the 2D HO basis functions. With the LFWFs for $\pi^+$ and $K^+$ obtained this way, the PDF for the valence quark is given by~\cite{Li:2017mlw} 
	\begin{equation}
	f(x)=\sum_{rs}\int \dfrac{d\overrightarrow{k}^\perp}{(2\pi)^2} \dfrac{\psi^*_{rs}(x,\overrightarrow{k}^\perp)\,\psi_{rs}(x,\overrightarrow{k}^\perp)}{4\pi\,x(1-x)},\label{eq:PDF_valence_psi}
	\end{equation}
while the PDF for the valence antiquark is given by ${f(1-x)}$. 
In our basis representation of Eq.~\eqref{eq:psi_rs_basis_expansions}, the transverse integrals in  Eq.~\eqref{eq:PDF_valence_psi} can be evaluated analytically using the orthonormal property of the 2D HO functions. 

We notice that finite $L_{\mathrm{max}}$ leads to a basis artifact: oscillations in the obtained PDFs. The amplitudes of such oscillations diminish with increasing $L_{\mathrm{max}}$. Therefore we fit the PDFs at different $L_{\mathrm{max}}\in \{8,\,12,\,16,\,20,\,24,\,28,\,32\}$ using a smooth parameterized form ${f(x)=x^a (1-x)^b/B(a+1,b+1)}$.
Here ${B(a+1,b+1)}$ is the Euler Beta function that ensures the normalization of the PDF. We then fit the $L_{\mathrm{max}}$ dependence of these fitting parameters $(a,\,b)$ by quadratic functions on $L^{-1}_{\mathrm{max}}$ and extrapolate to $L_{\mathrm{max}}\rightarrow +\infty$. The resulting parameters are ${a=b=0.5961}$ for the pion, while ${a=0.6337}$ and ${b=0.8546}$ for the kaon. 

We now have our PDFs for the pion and the kaon at scales relevant to constituent quark masses which are several hundred $\mathrm{MeV}$. At the model scales, both PDFs for the valence quark and the valence antiquark are normalized to $1$:
\begin{equation}
\int_{0}^{1}f(x)\,dx = \int_{0}^{1}f(1-x)\,dx=1.\label{eq:initial_PDF_normalization}
\end{equation}
Meanwhile, we have the following momentum sum rule: 
\begin{equation}
\int_{0}^{1}x\,f(x)\,dx +\int_{0}^{1}x\,f(1-x)\,dx=1,\label{eq:momentum_sum}
\end{equation}
which is a consequence of Eq.~\eqref{eq:initial_PDF_normalization}. 
Equation~\eqref{eq:momentum_sum} states that the valence quark and antiquark together carry the entire light front momentum of the meson, as is appropriate to a low-resolution model. 
\section{The QCD evolution of PDFs}
\begin{figure}
	\begin{center}
		\includegraphics[width=\linewidth]{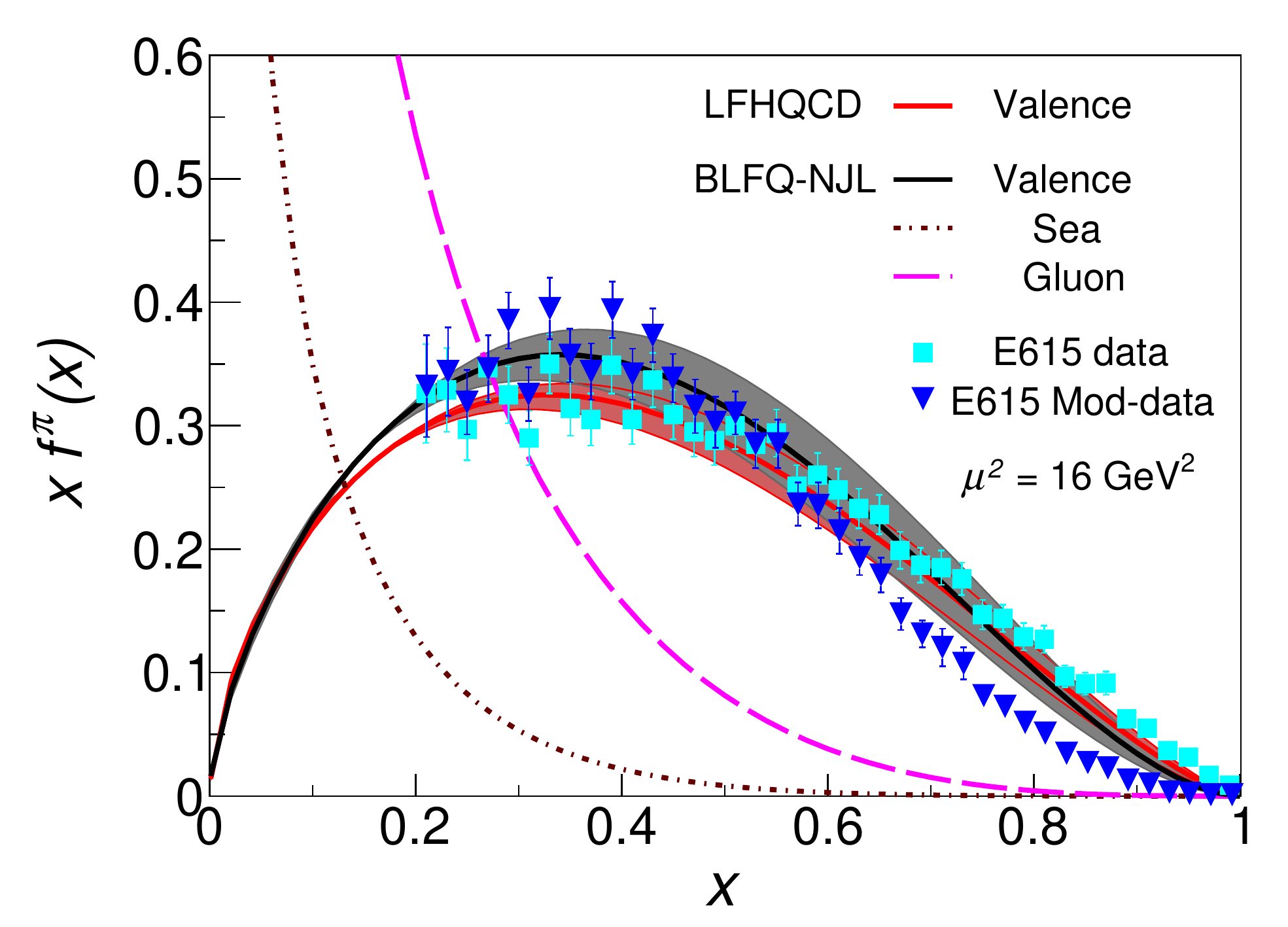}
		\caption{$xf^\pi(x)$ as a function of $x$ for the pion. The black bands are BLFQ-NJL results evolved from the initial scale $(0.240\pm 0.024~\mathrm{GeV}^2)$ using the NNLO DGLAP equations to the experimental scale of $16~\mathrm{GeV}^2$. The brown dot-dashed line and the pink long-dashed line represent our sea quark and gluon distributions, respectively, at the experimental scale using the same approach without uncertainties in our model scale. While the red band corresponds to light front holographic QCD predictions~\cite{deTeramond:2018ecg}. Results are compared with the original analysis of the FNAL-E615 experiment data~\cite{Conway:1989fs} and with its reanalysis (E615 Mod-data)~\cite{Aicher:2010cb}.}
		\label{fpionE615}
	\end{center}
\end{figure}
Next, we adopt the next-to-next-to-leading order (NNLO) Dokshitzer-Gribov-Lipatov-Altarelli-Parisi (DGLAP) equations~\cite{Dokshitzer:1977sg,Gribov:1972ri,Altarelli:1977zs} of QCD, to evolve our PDFs from our model scales, defined as $\mu_{0}^2$, to higher scales $\mu^2$ needed for the comparison with experiment. The scale evolution allows quarks to emit and absorb gluons, with the emitted gluons allowed to create quark-antiquark pairs as well as additional gluons~\cite{peskin1995introduction}. In this picture, the higher scale reveals the sea quark and gluon components of the constituent quarks through QCD. 

Explicitly we evolve our initial PDFs from the BLFQ-NJL model for the mesons to the relevant experimental scales ${\mu^2=16~\mathrm{GeV}^2}$ and ${\mu^2=20~\mathrm{GeV}^2}$ using the Higher Order Perturbative Parton Evolution toolkit to numerically solve the NNLO DGLAP equations~\cite{Salam:2008qg}. We determine ${\mu_{0\pi}^2=0.240\pm0.024~\rm{GeV}^2}$ for the pion and ${\mu_{0K}^2=0.246\pm0.024~\rm{GeV}^2}$ for the kaon by requiring the results after QCD evolution to fit both the pion PDF data from the FNAL-E615 experiment~\cite{Conway:1989fs} and the ratio $u^K_{v}/u^\pi_{v}$ data from the CERN-NA3 experiment \cite{Badier:1983mj}. The value of $\chi^2$ per degree of freedom (d.o.f.) for the fit of the pion PDF is 3.64, whereas for the ratio $u^K_{v}/u^\pi_{v}$ the value of $\chi^2/$d.o.f. is 0.50. We estimate a $10\%$ uncertainty in the initial scales. We also note that the best-fit initial scales increase $17\%$ when we advance the DGLAP equations from NLO to NNLO, with reduced $\chi^2~/~$(d.o.f.) and qualitatively comparable fitted PDFs. 

In Fig.~\ref{fpionE615}, we show our result for the valence quark PDF of the pion, where we compare the valence quark distribution after QCD evolution with the data from the FNAL-E615 experiment for the pion-nucleus-induced DY process~\cite{Conway:1989fs}. The error bands in our evolved valence quark distributions are due to the spread in the initial scale $\mu_{0\pi}^2=0.240\pm 0.024$ GeV$^2$. Our pion valence PDF falls off as $(1-x)^{1.44}$, favoring the slower falloff in the large $x$ region of the original analysis of the FNAL-E615 data~\cite{Conway:1989fs}. Our results differ from others in the same large $x$ region: Refs.~\cite{Chen:2016sno,Aicher:2010cb} favor the $(1-x)^2$ perturbative QCD falloff, while Ref.~\cite{deTeramond:2018ecg} supports a softer falloff of $(1-x)^{1.51}$. 

Looking further into the approach of Ref.~\cite{deTeramond:2018ecg}, which is based on the light front holographic QCD of Ref.~\cite{Brodsky:2014yha}, sea quark contributions to the pion PDFs were calculated using a nonvanishing $\vert q\overline{q}q\overline{q}\rangle$ Fock sector at their model scale $\mu_0^2=(1.12 \pm 0.32)$ GeV$^2$. At this scale, in their model the valence quarks carry $54\%$ of the pion's momentum, close to our model prediction of $57\%$. However, we note that there are significant differences between our model and theirs in how the remaining fraction of the pion momentum is distributed. Specifically, at this scale our model has $35\%$ of the total pion momentum in the gluons, while the corresponding contribution in  Ref.~\cite{deTeramond:2018ecg} is zero. 

\begin{figure}
	\begin{center}
		\includegraphics[width=\linewidth]{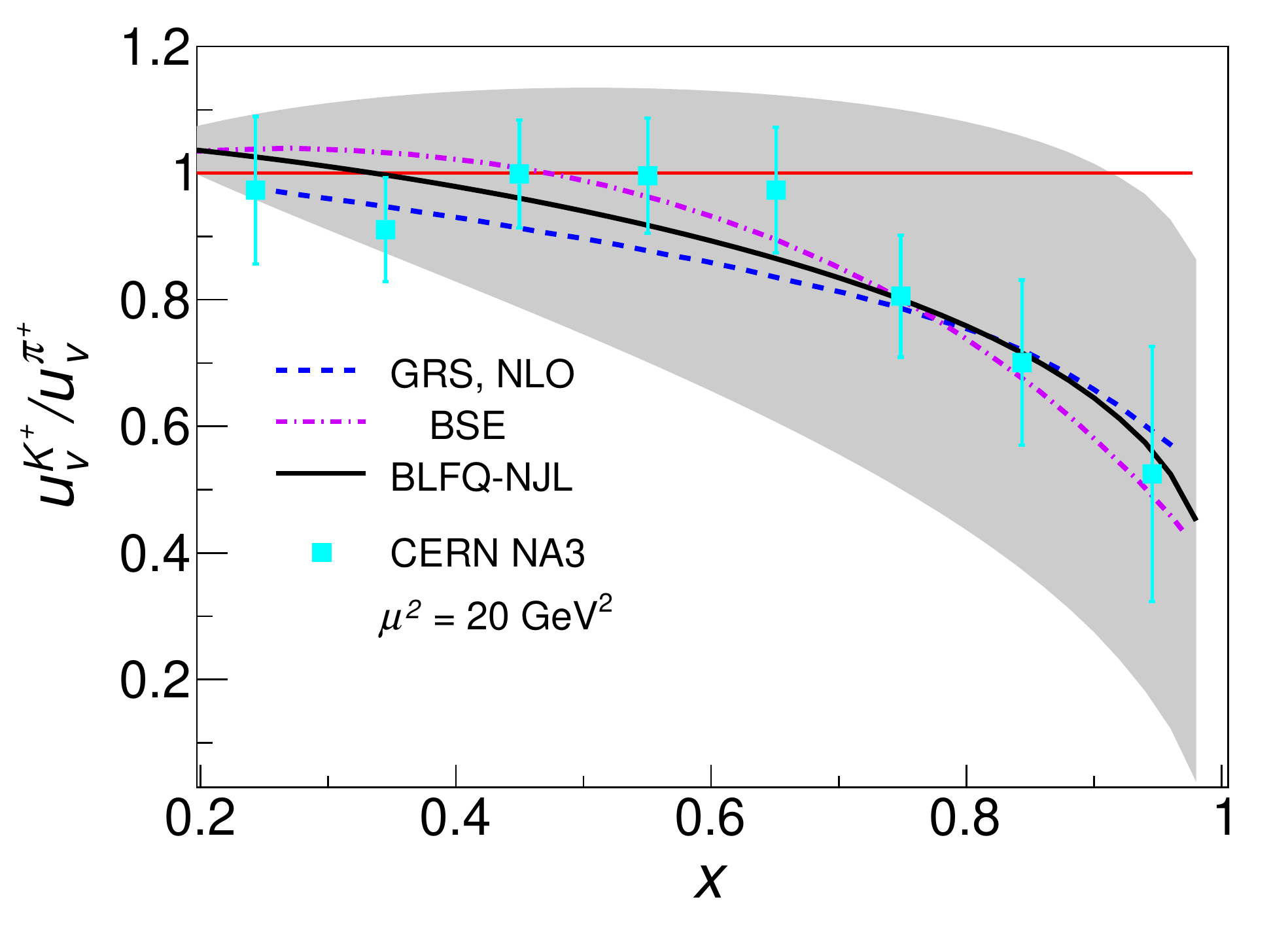}
		\caption{The ratio of the $u$ quark PDF in the kaon to that in the pion. The gray error band corresponds to the sum of relative errors due to the QCD evolution from the initial scale $\mu_{0\pi}^2=0.240\pm 0.024$ GeV$^2$  in the pion and ${\mu_{0K}^2=0.246\pm0.024~\rm{GeV}^2}$ in the kaon PDFs as the relative error for this ratio. The data are taken from the CERN-NA3 experiment~\cite{Badier:1983mj}. Results are compared with the NLO Glück-Reya-Stratmann (GRS) model~\cite{Gluck:1997ww} and the BSE approach~\cite{Nguyen:2011jy}.}
		\label{kaonpdf}
	\end{center}
\end{figure}
\begin{figure}[b]
	\begin{center}
		\includegraphics[width=\linewidth]{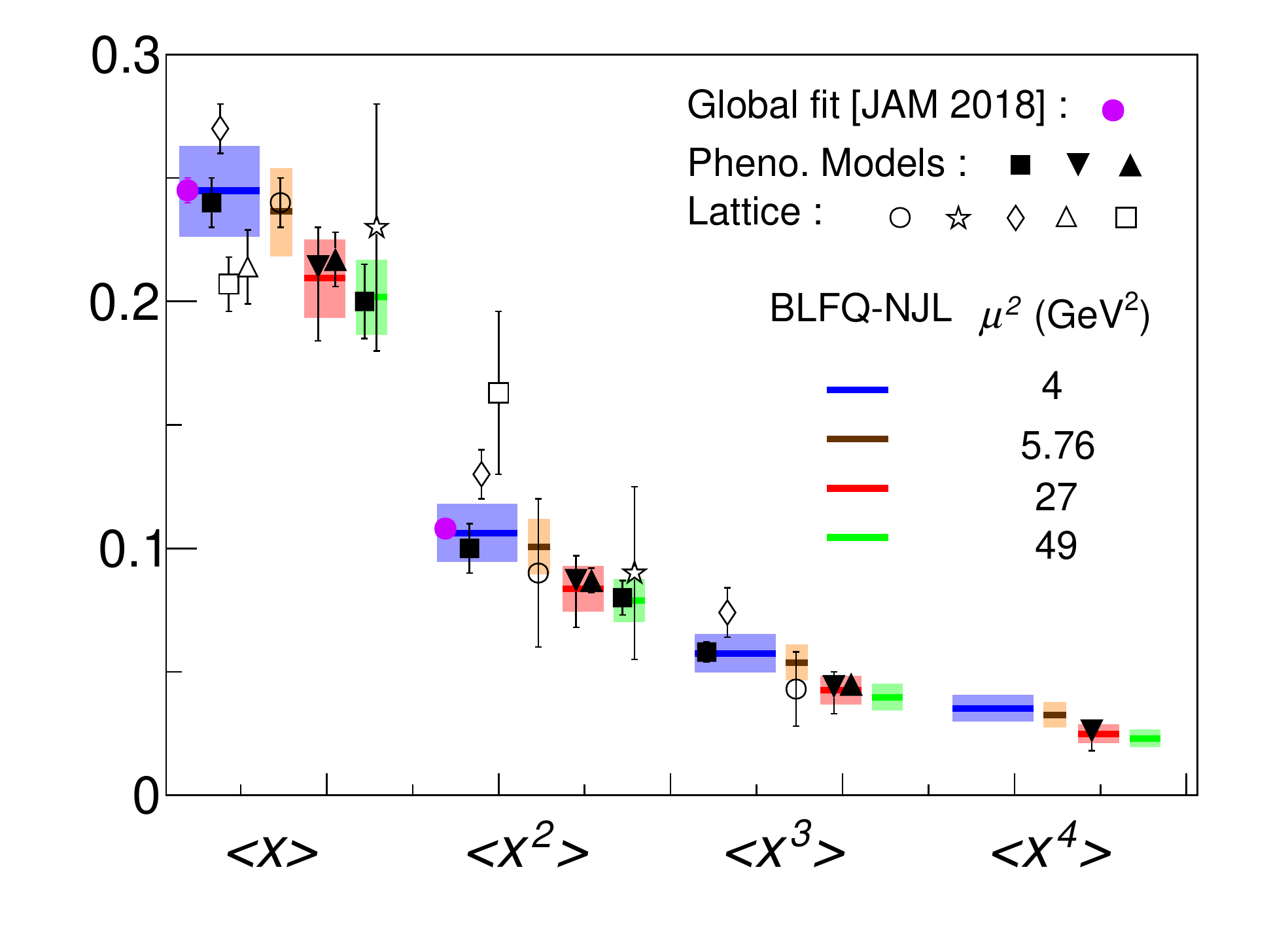}
		\caption{Comparison of the lowest four moments of valence quark distributions in the pion at four scales. Horizontal bands represent the BLFQ-NJL results including the uncertainty of the initial scale and are compared with the global fit to data by the JAM Collaboration~\cite{Barry:2018ort}, with lattice QCD results in Refs.~\cite{Brommel:2006zz,Martinelli:1987bh,Detmold:2003tm,Abdel-Rehim:2015owa,Oehm:2018jvm}, and with phenomenological models in Refs.~\cite{Nam:2012vm,Sutton:1991ay,Wijesooriya:2005ir} at different scales.}
		\label{moment}
	\end{center}
\end{figure}
We present the ratio of the $u$ quark distribution in the kaon to that in the pion in Fig.~\ref{kaonpdf}. We observe that at $\mu^2=20~\mathrm{GeV}^2$ our result for $u_{{v}}^{K^+}/u_{{v}}^{\pi^+}$, which is used to help determine the initial scale of the kaon PDF, is in good agreement, considering the current uncertainties, with the data from CERN-NA3 experiment~\cite{Badier:1983mj} as well as with the next-to-leading-order quark model (GRS, NLO)~\cite{Gluck:1997ww} and the Bethe-Salpeter equation (BSE) approach~\cite{Nguyen:2011jy}. One notices that the ratio decreases as $x$ increases. This phenomena can be understood from the valence quark PDFs of the kaon and the pion evolved to $\mu^2=20~\mathrm{GeV}^2$. Specifically, the antistrange ($\overline{s}$) quark is more likely than the ${u}$ quark to carry a large momentum fraction in the kaon, while the pion structure is symmetric in the antidown ($\overline{d}$) and ${u}$ quarks. We find additionally that at this experimental scale the $u$ quark PDF in the kaon falls off at large $x$ as $(1-x)^{1.60}$, in contrast to $(1-x)^{1.49}$ in the pion. On the other hand, at large $x$, the $\overline{s}$ quark PDF in the kaon falls off as $(1-x)^{1.32}$. Such differences among these PDFs are attributable to differences in the constituent quark mass~\cite{Jia:2018ary} propagated through the QCD evolution. 

To further compare the BLFQ-NJL model with experiment and with other models, we evaluate the four lowest  nontrivial moments of the valence quark PDF for the pion. In Fig.~\ref{moment}, we show these results at different $\mu^2$ and compare with the global fit to data~\cite{Barry:2018ort}, lattice QCD~\cite{Brommel:2006zz,Martinelli:1987bh,Detmold:2003tm,Abdel-Rehim:2015owa,Oehm:2018jvm}, and phenomenological models~\cite{Nam:2012vm,Sutton:1991ay,Wijesooriya:2005ir}. Figure~\ref{moment} shows that our predictions are in good agreement with Refs.~\cite{Barry:2018ort,Nam:2012vm,Sutton:1991ay,Wijesooriya:2005ir,Detmold:2003tm,Martinelli:1987bh}.

\begin{figure}
	\begin{center}
		\includegraphics[width=\linewidth]{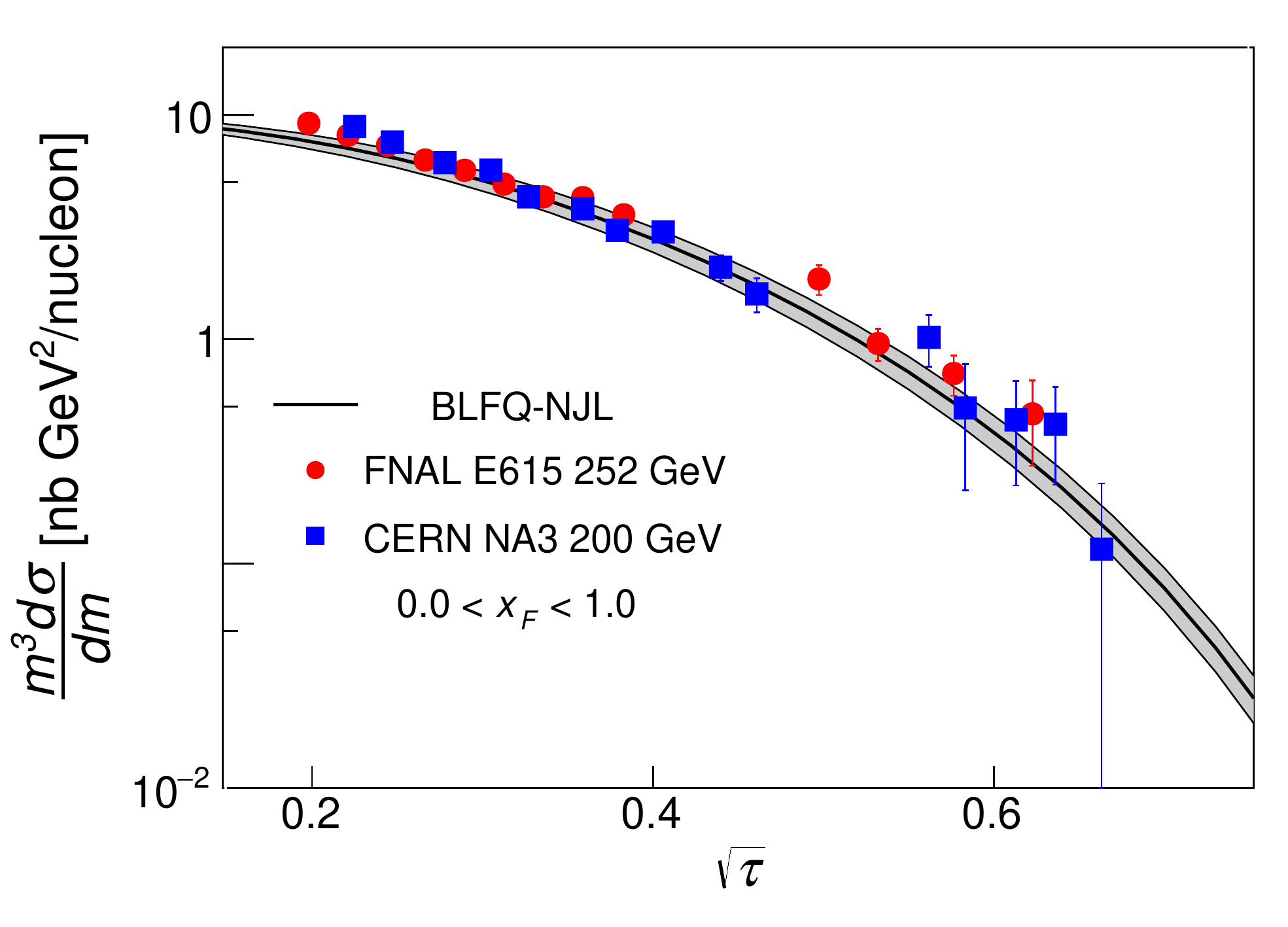}
		\includegraphics[width=\linewidth]{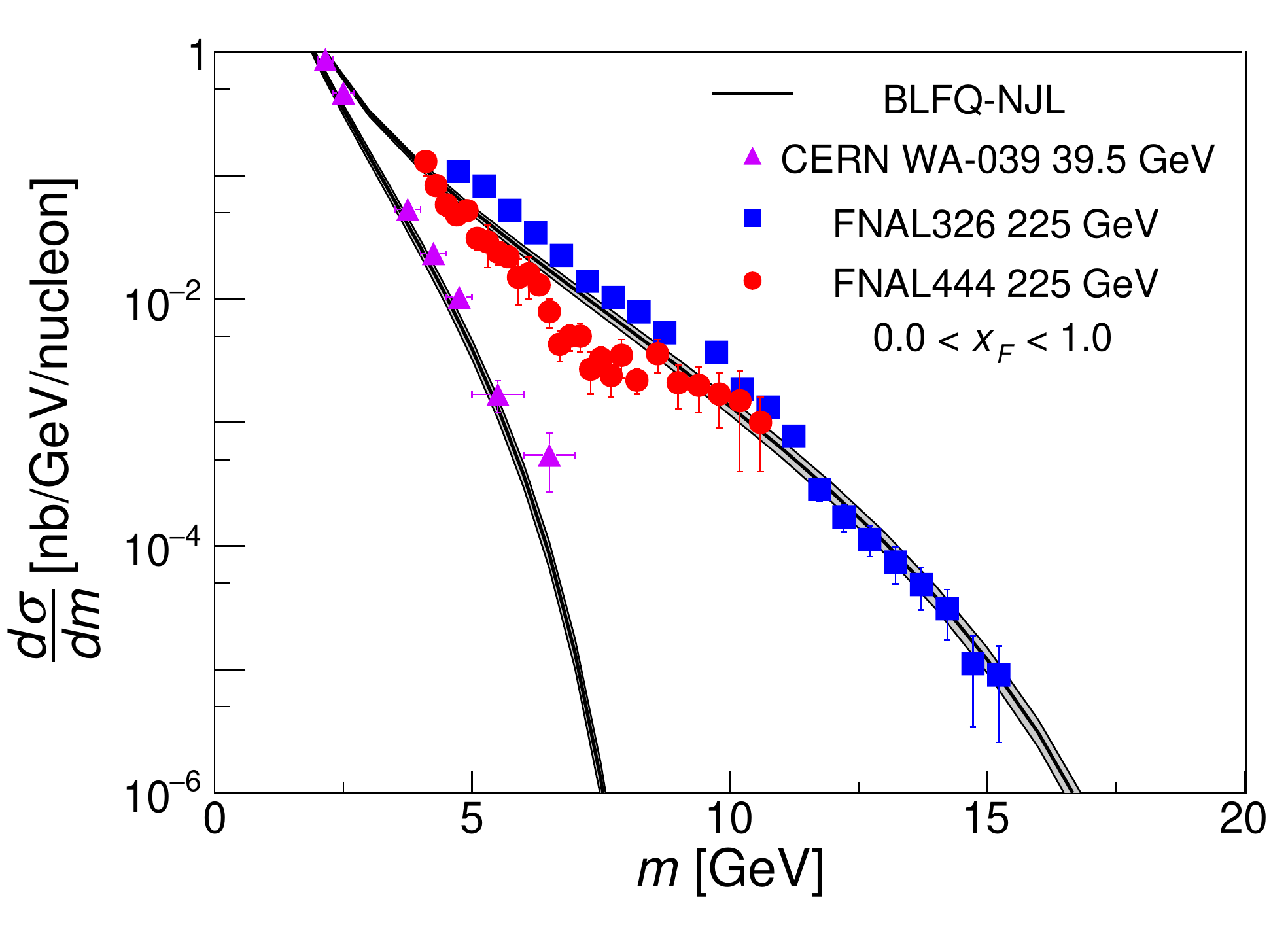}
		\caption{$m^3 d{\bf \sigma}/dm$ for the $\pi^-$-nucleus DY process as a function of $\sqrt{\tau}$ and $d{\bf \sigma}/dm$ as a function of $m$. The data are taken from Refs.~\cite{Conway:1989fs,Badier:1983mj,Stirling:1993gc,Greenlee:1985gd,Anderson:1979tt,Corden:1980xf}.
		The FNAL444 and the CERN NA3 data correspond to a carbon target and a platinum target, respectively. The FNAL326, the CERN WA-039, and the FNAL-E615 data all use a tungsten target.}
		\label{cross1}
	\end{center}
\end{figure}

The kinematics of the pion-nucleus-induced DY process are described by the invariant mass of the produced lepton pair $m$, center of mass energy square $s$ of the colliding systems, the Feynman variable ${x_{F}=x_1-x_2}$ (difference of the light front momenta of the annihilating quark and antiquark), and ${\tau\equiv m^2/s=x_1x_2}$~\cite{Pasquini:2014ppa}. In the leading order of QCD, the cross section for this process is given by~\cite{Drell:1970wh,McGaughey:1994dx}
\begin{equation}
\frac{m^3 d^2{\bf \sigma}}{dm\, dx_{\mathrm{F}}}=\frac{8\pi\alpha^2}{9}\frac{x_1x_2}{x_1+x_2}\sum_a e_a^2f^{\pi^\pm}_a(x_1){f}^N_{\bar a}(x_2),\label{crosseq}
\end{equation}
where $\alpha$ is the coupling constant of quantum electrodynamics. The summation in Eq.~\eqref{crosseq} runs over different quark flavors, with $e_a$ being their charges in units of the elementary charge. Here, we use our pion PDFs in conjunction with the NNLO ``MSTW~2008'' nucleon PDFs~\cite{Martin:2009iq}. We ignore the European Muon Collaboration effect~\cite{Aubert:1983xm} and treat the target nucleus as a collection of free nucleons. The nucleon and the pion PDFs are then evolved to the experimental scale ${\mu^2=16~\mathrm{GeV}^2}$. After integrating out the $x_{F}$ dependence of the cross section to yield $m^3 d{\bf \sigma}/dm$, we obtain our results plotted as functions of $\sqrt{\tau}$ in the upper panel of Fig.~\ref{cross1} and compared with the CERN-NA3 and FNAL-E615 experiments. In the lower panel of Fig.~\ref{cross1}, we illustrate the cross section $d{\bf \sigma}/dm$ as a function of $m$ and compare with the FNAL-326~\cite{Greenlee:1985gd} and the FNAL-444 experiments~\cite{Anderson:1979tt} with $225~\mathrm{GeV}$ pions. In addition, we compare our results with the data of CERN-WA-039 experiment with $39.5~\mathrm{GeV}$ pions~\cite{Corden:1980xf}. All BLFQ-NJL results in Fig.~\ref{cross1} are in reasonable agreement with experiment. Here, we have selected sample experimental cases over a wide kinematic range for validating the BLFQ-NJL model. We note that our approach yields comparable agreement with results from other experimental setups~\cite{Conway:1989fs,Badier:1983mj,Stirling:1993gc,Barate:1979da} as will be detailed elsewhere \cite{Lan:2019zz}. 
\section{Conclusion}
We present a model for the pion and the kaon that unifies their properties from the low-resolution constituent quark picture to high-resolution experiments. Specifically, we begin with an effective light front Hamiltonian incorporating confinement and chiral dynamics for a valence quark-antiquark pair suitable for low-resolution properties. Using basis light front quantization~\cite{Jia:2018ary}, the parameters in this Hamiltonian were adjusted to reproduce the experimental mass spectrum and charge radii of the light mesons~\cite{Jia:2018ary}. The light front wave functions obtained as the eigenvectors of this Hamiltonian were then used to generate the initial PDFs. The corresponding PDFs at higher experimental scales have been computed based on the NNLO DGLAP equations. 

The initial low-resolution scales are the only adjustable parameters involved in QCD scale evolution, and we obtain them by simultaneously fitting the FNAL-E615~\cite{Conway:1989fs} and the CERN-NA3 experimental data~\cite{Badier:1983mj}. We then find leading moments of the pion PDF over a range of scales agree with results from the global fit in Ref.~\cite{Barry:2018ort}, the results from lattice QCD~\cite{Detmold:2003tm,Martinelli:1987bh}, and the results from phenomenological quark models~\cite{Nam:2012vm,Sutton:1991ay,Wijesooriya:2005ir}. We have also calculated the cross sections of the pion-nucleus-induced Drell-Yan process and have obtained good agreement with available data~\cite{Conway:1989fs,Badier:1983mj,Stirling:1993gc,Barate:1979da,Greenlee:1985gd,Anderson:1979tt,Corden:1980xf}. We note that the valence PDFs at the corresponding experimental scales for the pion and the kaon based on the NNLO DGLAP equations are almost unchanged from the NLO results, while the fitted initial scales based on the NNLO equations are $17\%$ higher than those obtained at NLO. These favorable results confirm the robust character of the BLFQ-NJL model which includes QCD evolution, motivating the application of analogous effective Hamiltonians to the baryons with subsequent scale evolution.
\begin{acknowledgments}
We thank  J. Huston, W. Zhu, J. Gao, X. Chen, T. Liu, and R. Wang for many insightful discussions. C. M. is supported by the China Postdoctoral Science Foundation (CPSF) under the Grant No.~2017M623279 and the National Natural Science Foundation of China (NSFC) under the Grant No. 11850410436. This work of X. Z. is supported by new faculty startup funding by the Institute of Modern Physics, Chinese Academy of Sciences under the Grant No.~Y632030YRC.  S. J. and J. P. V. are supported  by the Department of Energy under Grants No.~DE-FG02-87ER40371 and No.~DE-SC0018223 (SciDAC4/NUCLEI).
\end{acknowledgments}
\bibliographystyle{apsrev4-1}
\bibliography{BLFQ_NJL_prl_bib}
\end{document}